\documentclass[preprint,aps]{revtex4}

\usepackage{graphicx}
\usepackage{dcolumn}
\usepackage{bm}

\begin{document}

\preprint{}

\title{Parity, Charge Conjugation and $SU(3)$ Constraints on\\
Threshold Enhancement in $J/\psi$ decays into $\gamma p\bar p$ and
$K p \bar\Lambda$ }

\author{Xiao-Gang He}

 \altaffiliation[Also at]{ Department of Physics, Peking
 University;
 On leave of absence from Taiwan University, Taipei}
 \email{hexg@phys.ntu.edu.tw}
\author{Xue-Qian Li}%
 \email{lixq@nankai.edu.cn}
\affiliation{%
Department of Physics, Nankai University, Tianjin, China\\
}%
\author{J.P. Ma}%
 \email{majp@itp.ac.cn}
\affiliation{%
Institute of Theoretical Physics, Academia Sinica, Beijing, China\\
}%

\date{\today}

\begin{abstract}
We study the threshold enhancement effects of baryon-anti-baryon
systems in $J/\psi$ decays at BES using parity $P$, charge
conjugation $C$, and flavor $SU(3)$ symmetries. The $P$ and $C$
symmetries restrict the $p\bar p$ in $J/\psi \to \gamma p\bar p$
to be in a state with $C=+1$, while the $p\bar p$ in $J/\psi \to
\pi^0 p\bar p$ to be with $C=-1$. Combining the $C$ and $P$
symmetries with flavor $SU(3)$ symmetry, i.e. the CPS symmetry, we
find that the $\Lambda \bar p$ system cannot be in $0^+$ and $0^-$
states in $J/\psi \to K^+ \Lambda \bar p$. We provide a consistent
explanation of observation and non-observation of near threshold
enhancements in $J/\psi \to K^+ \Lambda \bar p$ and $J/\psi \to
\pi^0 p \bar p$, respectively. We also find that near baryon pair
threshold enhancement can happen in several channels for $J/\psi$
decays and  can be several times larger than that observed in
$J/\psi \to K^+ \Lambda \bar p$ in some channels.
\end{abstract}

\pacs{Valid PACS appear here}
\maketitle

Recently the BES collaboration has observed enhancements in
$J/\psi \to\gamma p \bar p$\cite{bes1} and $J/\psi \to K p \bar
\Lambda$\cite{bes2} with the invariant masses of the $p\bar p$ and
$p \bar \Lambda$ near their thresholds, whereas no enhancement in
$J/\psi \to \pi^0 p\bar p$\cite{bes1} is seen. Similar enhancement
effects have also been observed in $B^+\to K^+ p \bar p$ and $\bar
B^0 \to D^0 p\bar p$\cite{belle}. There are no known states
corresponding to the required resonant masses. These events are
quite anomalous.
\par
These anomalous events have stimulated a number of theoretical
speculations\cite{e1,e2,ea,e3,e4,e5,bugg,zhu}. In \cite{e1} the
enhancement is explained by an $^1S_0$ bound state of $p\bar p$.
The binding energy and lifetime of such a state and also a
$P$-wave state is studied within the linear $\sigma$
model\cite{e2}. In \cite{ea} scalar glueball mechanism is invoked.
Several other mechanisms are also discussed. The effect of final
state interactions through one pion exchange is studied in
\cite{e3} and it is found that the enhancement can be partly
reproduced. In \cite{e4} LEAR data are used to fix scattering
lengthes of $p\bar p$ scattering near the threshold and with these
fixed scattering lengthes an enhancement near the threshold of
$J/\psi\to\gamma p \bar p$ can be produced and also an explanation
of no enhancement in $J/\psi\to\pi^0 p\bar p$ can be given. In
\cite{e5} a $ 0^{-+}$ baryonium close to the threshold is
predicted with a Skyrmion-type potential. In \cite{bugg} it is
suggested that the enhancement in $J/\psi \to \gamma p\bar p$ may
be fitted as a cusp. All of these works can more or less explain
the enhancement in $\gamma p\bar p$, but a decisive explanation is
still needed. In \cite{zhu} possible quantum numbers of the
possible resonant state in $\gamma p\bar p$ and its possible decay
modes are discussed.
\par
Although the dynamical mechanism responsible for these anomalous
events is not understood in detail, the strong and/or
electromagnetic interaction should be responsible for the decays
and the enhancement near the threshold. The decays must respect
parity $P$ and charge conjugation $C$ symmetries. These symmetries
can provide important information about the enhancement effects.
We will perform such an analysis for the decays of a $J/\psi$ into
$\gamma p\bar p$, $\pi^0 p\bar p$. We further carry out a more
general analysis for the decays of a $J/\psi$ into $\gamma B' \bar
B,\;MB'\bar B$, where $M$ stands for an octet meson in the flavor
$SU(3)$ symmetry, and $B'(\bar B)$ stands for an octet
baryon(anti-baryon) trying to use $SU(3)$ symmetry to relate and
to constrain $\pi^0 p\bar p$ and $K^+ \Lambda \bar p$ and provide
new predictions for processes involving other baryons and mesons,
i.e., $J/\psi \to \gamma B' \bar B$ and $J/\psi \to M B' \bar B$.

We first consider possible constraints on decay amplitudes of
$J/\psi$ from $P$ and $C$ symmetries. We take $J/\psi\to \gamma
p\bar p$ as an example for some details of our analysis. For
convenience of identifying quantum numbers of the baryon pairs
near their thresholds, we study the decay amplitude in the
rest-frame of $p\bar p$. We have
\begin{equation}
J/\psi (\vec p, \vec\varepsilon_J)  \to \gamma(\vec p,
\vec\varepsilon ) + p(\vec q) +\bar p(-\vec q),
\end{equation}
where $\vec \varepsilon_J$ and $\vec \varepsilon$ are 3-vectors
for the polarizations of the $J/\psi$ and photon, respectively.
The particle 3-momenta are indicated in brackets. The decay
amplitude can be written with standard Dirac spinors as:
\begin{equation}
{\mathcal T}(J/\psi\to\gamma p\bar p) = \bar u(\vec q) A(\vec\varepsilon_J,
\vec\varepsilon, \vec p, \vec q) v(-\vec q),
\end{equation}
with $A$ being a $4\times 4$ matrix. For our purpose it is
convenient to work with two-component spinors using $\psi^\dagger$
for the proton and $\chi $ for the anti-proton.
\par
The Dirac spinor $\bar u(\vec q )$($v(-\vec q)$) can be expressed
with $\psi^\dagger $ ($\chi$), $\vec q$ and $\vec\sigma$, where
$\sigma^i (i=1,2,3)$ are Pauli matrices. The decay amplitude can
be then written with these two-component spinors as:
\begin{equation}
{\mathcal T}(J/\psi\to\gamma p\bar p) = \psi^\dagger
\left ( F_{\gamma 0}(\vec p, \vec q) +\vec\sigma \cdot \vec F_{\gamma\sigma} (\vec
p, \vec q) \right ) \chi.
\end{equation}
The form factors $F_{\gamma 0}(\vec p, \vec q)$ and $\vec
F_{\gamma\sigma} (\vec p, \vec q)$ also depend on polarization
vectors. Near the threshold of the $p\bar p$ system, one can
expand the form factors in $|\vec q|$. In the expansion one
identifies the $p\bar p$ system with quantum numbers according to
the total spin $S$, the orbital angular momentum $L$ and the total
angular momentum $J=L+S$, in short $^{2S+1} L_J$. The $P$ and $C$
eigenvalues of the system are given by $(-1)^{L+1}$ and
$(-1)^{L+S}$, respectively.
\par
The form factor $F_{\gamma 0}$ can be
written as:
\begin{equation}
F_{\gamma 0} = \varepsilon_J^{i}\varepsilon^{j} T_0^{ij}(\vec p, \vec q),
\end{equation}
while the tensor $T_0^{ij}$ can be decomposed with the following 9
tensors of rank 2 because of rotation covariance: $\delta_{ij},\
\varepsilon_{ijk} p^k,\   \varepsilon_{ijk} q^k,\
 \varepsilon_{ijk} n^k,
 p^{ \{ i} p^{j\}}$,
$q^{ \{i} q^{j\} }, \ p^{ \{i} q^{j\}},\
 p^{ \{i} n ^{j\}},\   n^{ \{i} q^{j\}}$,
with $\vec n = \vec p \times \vec q$. The notation $S^{\{ ij\}}$
means that $S$ is symmetric and trace-less. Parity conservation
eliminates terms proportional to $\delta_{ij},\ \varepsilon_{ijk}
n^k,
 p^{ \{ i} p^{j\}},\ q^{ \{i} q^{j\} }, \ p^{ \{i} q^{j\}}$ in $T_0^{ij}$.
Hence the tensor is written as:
\begin{eqnarray}
&&T_0^{ij}(\vec p, \vec q) = \varepsilon_{ijk} p^k b_1(m_{p\bar
p}, \nu )
          +\varepsilon_{ijk} q^k b_2(m_{p\bar p}, \nu )\nonumber\\
          &&+p^{ \{i} n ^{j\}} b_3 (m_{p\bar p}, \nu )
          + q^{ \{i} n ^{j\}} b_4 (m_{p\bar p}, \nu ),
\end{eqnarray}
with $\nu = \vec p\cdot \vec q$. It should be noted that
$b_i(i=1,2,3,4)$ are Lorentz invariant form factors, they depend
on the invariant mass $m_{p\bar p}$ of the $p\bar p$ system and
$\nu$, the later can be expressed with $m_{p\bar p}$ and the
invariant mass of the $p\gamma$ system. The symmetry of charge
conjugation gives constraints of these form factors: $
b_1(m_{p\bar p}, \nu) = b_1(m_{p\bar p}, -\nu), \ b_2(m_{p\bar p},
\nu) = -b_2(m_{p\bar p}, -\nu)$, $ b_3(m_{p\bar p}, \nu)=
-b_3(m_{p\bar p}, -\nu), \ \ b_4(m_{p\bar p}, \nu) = b_4(m_{p\bar
p}, -\nu)$. Near the threshold, i.e., $\vec q \to 0$, one can
expand these form factors in $\nu$.
This expansion is equivalent to the standard partial wave
analysis. Throughout this work we will only consider contributions
up to $D$-wave states because contributions from states with $L>2$
are suppressed by at least $|\vec q|^3$ which are small compared
with baryon masses near the threshold. With this approximation
$F_{\gamma 0} $ takes the form:
\begin{eqnarray}
&&F_{\gamma 0} = (\vec\varepsilon \times \vec\varepsilon_J )
\cdot\vec p {\mathcal S}_{\gamma}
               +\left (q^i q^j - \frac{1}{3}\vert \vec q\vert^2 \delta_{ij} \right )
               \nonumber\\
               && \cdot \Big [ (\vec\varepsilon \times \vec\varepsilon_J ) \cdot\vec p p^i p^j
            {\mathcal D}_{1\gamma}
            +(\vec\varepsilon\times\vec\varepsilon_J )^i p^j {\mathcal
           D}_{2\gamma}\nonumber\\
           &&+\vec\varepsilon_J\cdot\vec p (\vec \varepsilon \times \vec p)^i p^j
             {\mathcal D} _{3\gamma}
             +(\varepsilon^i (\vec\varepsilon_J\times \vec p)^j\nonumber\\&& +
             +\varepsilon_J^i (\vec\varepsilon\times \vec p)^j) {\mathcal D}_{4\gamma}
     \Big ] +\cdots ,\label{gamma1}
\end{eqnarray}
where we have used $\vec p\cdot\vec \varepsilon =0$ for the
photon. All the form factors ${\mathcal S}_\gamma$, ${\mathcal
D}_{i\gamma}(i=1,2,3,4)$ only depend on $m_{p\bar p}$.

\par
One can also make a general decomposition for $\vec F_{\gamma
\sigma}$. It has the form:
\begin{equation}
F_{\gamma\sigma} ^ i = \varepsilon_J^j \varepsilon^{k}
T_\sigma^{ijk} (\vec p, \vec q).
\end{equation}
The constraints for the tensors are: $ T_\sigma ^{ijk} (\vec p,
\vec q) = -T_\sigma^{ijk} (-\vec p, -\vec q)$ from parity, and $
T_\sigma^{ijk} (\vec p, \vec q) = -T_\sigma^{ijk} (\vec p, -\vec
q)$ from charge\ conjugation. From the above analysis one can
easily see that $\vec F_{\gamma\sigma}$ represents the
contributions of the states with $L=1,3,\cdots$, i.e., $P$-wave
states, $F$-wave states,\ $\cdots$. The general expression is
complicated since the tensor $T^{ijk}$ is of rank 3. Using the
fact $\vec\varepsilon\cdot\vec p =0$, we have for the leading
non-zero terms with $l = 1$:
\begin{eqnarray}
&&T_{\sigma}^{ijk} = \vec p\cdot\vec q \left [ p^i \delta_{jk}
{\mathcal P}_{1\gamma}
      + p^j\delta_{ik} {\mathcal P}_{2\gamma}
\right ] + p^i p^j q^k {\mathcal P}_{3\gamma}+ \cdots,\nonumber\\
\label{gamma2}
\end{eqnarray}
where ${\mathcal P}_{i\gamma}(i=1,2,3)$ are Lorentz form factors
depending on $m_{p\bar p}$.
\par
 The charge conjugation gives a constraint that $\vec
F_{\gamma\sigma}$ must {\it exactly} go to zero when $\vert \vec q
\vert \to 0$, thus, $\vec F_{\gamma\sigma}$ is not singular as
$\vert \vec q \vert \to 0$. It is well-known that Coulomb
interaction can cause a singularity in the amplitude when $\vert
\vec q \vert \to 0$. From our above discussion one sees that the
singularity can only appear in $F_{\gamma 0}$, i.e., in the
contribution with $S=0$.

\par
We now discuss the enhancement effect near the threshold observed
at BES\cite{bes1} using results obtained in the above. In
\cite{bes1} the effect is interpreted as an existence of a
resonant structure near the threshold. The resonant structure can
be an $S$-wave state or a $P$-wave state. If it is an $S$-wave
state, it must be the state with the quantum numbers $^1S_0$, or
$J^{PC}=0^{-+}$. If it is a $P$-wave state, it must be the state
with the quantum numbers $^3 P_{J}$, or $J^{PC}=J^{++}$ with
$J=0,1,2$. The effect of the resonant structure manifest itself
via $F_{\gamma 0}$, or $\vec F_{\sigma\gamma}$. Beside possible
enhancement effects due to the resonant structure near the
threshold, some other enhancement mechanisms should be in these
form factors for $\vert \vec q \vert \to 0$ in order to compensate
the suppression factor $\vert \vec q \vert \to 0$ from the
phase-space near the threshold. Coulomb interaction may do the
job\cite{e4}.

\par
In a similar way we can write the form factors for $J/\psi (\vec
p, \vec\varepsilon_J)  \to \pi^0(\vec p) + p(\vec q) +\bar p(-\vec
q)$. The amplitude  which respects $P$ and $C$ symmetries can be
written as
\begin{eqnarray}
&&{\mathcal T}(J/\psi\to\pi^0 p\bar p) =   \psi^\dagger \left (
F_{\pi 0}(\vec p, \vec q) +\vec\sigma \cdot \vec F_{\pi\sigma}
(\vec p, \vec q) \right ) \chi
\nonumber\\
   &&=\psi^\dagger
\Big \{  ( \vec\varepsilon_J \cdot \vec q {\mathcal P}_{1\pi} +
\vec\varepsilon_J \cdot \vec p \vec p \cdot \vec q {\mathcal
P}_{2\pi} ) +(\vec\sigma \times \vec\varepsilon_J ) \cdot\vec p
{\mathcal S}_{\pi}\nonumber\\
               &&+\Big (q^i q^j - \frac{1}{3}\vert \vec q\vert^2 \delta_{ij} \Big )
 \cdot \Big [ (\vec\sigma \times \vec\varepsilon_J ) \cdot\vec p p^i p^j
   {\mathcal D}_{1\pi}
 \nonumber\\
       &&    +(\vec\sigma\times\vec\varepsilon_J )^i p^j {\mathcal D}_{2\pi}
           +(\vec\sigma\cdot\vec p (\vec \varepsilon_J \times \vec p)^i
           p^j\nonumber\\
           &&+\vec\varepsilon_J\cdot\vec p (\vec \sigma \times \vec p)^i p^j {\mathcal D}_{3\pi}
           +(\sigma^i (\vec\varepsilon\times \vec p)^j\nonumber\\
             &&+\varepsilon_J^i (\vec\sigma\times \vec p)^j) {\mathcal D}_{4\pi}
     \Big ]
   \Big  \}  \chi  +\cdots,
\end{eqnarray}
where we have neglected contributions from states with $L>2$.
From the above a
possible resonant structure near the threshold for this decay
could only be the state with $S=0$ and $L=1$ and the quantum
numbers $J^{PC}=1^{+-}$, or with $S=1$ and $L=0,2$ and the quantum
numbers $J^{PC}=J^{--}$ where $J$ can be 1,2,3. Since the quantum
numbers here are different from those of the resonant structure in
$J/\psi\to\gamma p\bar p$, the observation of the resonant
structure in $J/\psi\to\gamma p\bar p$ does not necessarily
indicate the existence of a resonant structure in $J/\psi\to\pi^0
p\bar p$.

\par
Now we turn to the observed enhancement near the threshold in
$J/\psi \to K^+ \Lambda\bar p$\cite{bes2}. One needs to understand
why there is no enhancement in $\pi^0 p\bar p$ whereas there is
one in $K^+ \Lambda \bar p$. Because the final state $K^+ \Lambda
\bar p$ is not an eigenstate of the charge conjugation, we cannot
use the symmetry of charge conjugation to constrain the decay
amplitude in the same way as above. However, under $SU(3)$ the two
decays are related since the mesons $\pi^0$, $K^+$ and the baryons
$p$ and $\Lambda$ belong to octets of $SU(3)$. Therefore combining
$SU(3)$ with the $C$ and $P$ symmetries, one can obtain additional
constraints. This combined CPS\cite{soni} symmetry has been shown
to be extremely powerful in the lattice calculations of hadronic
matrix elements. One can expect that the CPS symmetry will set
further constraints on $J/\psi \to \pi^0 p\bar p$ and $J/\psi \to
K^+ \Lambda \bar p$ decays and their threshold enhancements.

The $J/\psi$ transforms as an $SU(3)$ singlet whereas $\pi^0$,
$K^+$, $p$ and $\Lambda$ transform as components of an $SU(3)$
octet. Using flavor $SU(3)$ symmetry one can write down relations
between different decay amplitudes for $J/\psi$ decays into a
pseudoscalar in the octet $M:(\pi^{\pm,0}, K^\pm, \bar K^0, K^0,
\eta)$ plus a baryon and an anti-baryon in the octet $B:
(\Sigma^{-,0,+}, p,n,\Xi^{-,0}, \Lambda)$.
The decay amplitudes respecting the $SU(3)$ symmetry depend on two
types of amplitudes $\tilde F$ and $\tilde D$ similar to
pseudoscalar nucleon couplings. Again we will use two-component
fields for baryons, hence the matrix elements in $B$ are two
component spinors. We can also decompose the $SU(3)$ fields as: $B
= {\sqrt{2}}\psi^{a\dagger}_B T^a$, $\bar B  = {
\sqrt{2}}\chi^{a}_{\bar B} T^a$, $M=\sqrt{2}M^a T^a$.  Here
$T^a$ is the $SU(3)$ generators normalized as
$Tr(T^aT^b)=\delta^{ab}/2$. The amplitude with the $SU(3)$
symmetry for $J/\psi (\vec p, \vec\varepsilon_J)  \to M (\vec p) +
B (\vec q) +\bar B (-\vec q)$ can be written as:
\begin{eqnarray}
&&{\mathcal T} (J/\psi\to MB\bar B) \nonumber\\&&= {\rm Tr} \left
[  ( \tilde D(\vec \varepsilon_J, \vec p, \vec q) \{M, B\} +
\tilde
F(\vec \varepsilon_J, \vec p,\vec q) [M, B]) \bar B \right ]\nonumber\\
&&= {\sqrt{2}}M^b \psi^{a\dagger}_{B} \left \{ d_{abc} \tilde
D(\vec \varepsilon_J, \vec p, \vec q)
  + if_{abc}\tilde F(\vec \varepsilon_J, \vec p,\vec q) \right \} \chi^c_{\bar B},
  \label{su3}\nonumber\\
\end{eqnarray}
where $f_{abc}$ and $d_{abc}$ are the $SU(3)$ structure constants.
$\tilde F$ and $\tilde D$ are $2\times 2$ matrices acting in the
spin space. Under charge conjugation the bilinear products of
baryon fields and the meson fields transforms as: $
\psi^{a\dagger}_{B} \chi^b_{\bar B} \to  c_b c_a
\psi ^{b\dagger}_{B} \chi^a_{\bar B}$, $ \psi^{a\dagger}_{B}
\vec\sigma \chi^b_{\bar B} \to -c_a c_b \psi^{b\dagger}_{
B}\vec\sigma \chi^a_{\bar B}$, $M^a \to c_a M^a $, where $
 c_a =  1,\; {\rm for}\ a=1,3,4,6,8$, $c_a = -1,\; {\rm for}\
 a=2,5,7$.
For the $SU(3)$ structure constants $f_{abc}$ and $d_{abc}$, we
have: $
 f_{abc} c_a c_b c_c = -f_{abc}$, $d_{abc} c_a c_b c_c = d_{abc}$.
Parity transformation is as usual.
\par
\par
Neglecting here all contributions with $L>1$, $\tilde F$ and
$\tilde D$ can be parameterized as:
\begin{eqnarray}
\tilde F &=&  \vec\varepsilon_J \cdot \vec q {\mathcal P}_{1F} +
\vec\varepsilon_J \cdot \vec p \vec p \cdot \vec q {\mathcal P}_{2F}
+(\vec\sigma \times \vec\varepsilon_J ) \cdot\vec p {\mathcal S}_{F}  +
\cdots
\nonumber\\
\tilde D &=&  \vec\varepsilon_J \cdot \vec q {\mathcal P}_{1D} +
\vec\varepsilon_J \cdot \vec p \vec p \cdot \vec q {\mathcal
P}_{2D} +(\vec\sigma \times \vec\varepsilon_J ) \cdot\vec p
{\mathcal S}_{D} +\cdots ,\nonumber\\
\label{sss}
\end{eqnarray}
where all form factors ${\mathcal P}_{(1,2)F}$, ${\mathcal
P}_{(1,2)D}$ and ${\mathcal S}_{F,D}$ depend only on the invariant
mass of $m_{B'\bar B}$. The general feature of the decay amplitude
is that if the state $B'\bar B$ is in the state with even $L$, it
must be in a spin triplet with $J^P =J^-$. If the state $B'\bar B$
is in the state with odd $L$, it must be in a spin singlet with
$J^P=L^+$.
From the general amplitude in
Eq.(\ref{su3}) we can find various decay amplitudes for different
decays, which are listed in Table I. In Table I all symbols for
baryons or anti-baryons stand for two- component spinors
respectively.
\par

\begin{table}[htb]
\caption{ $SU(3)$ amplitudes  for $J/\psi \to M B' \bar B$.
}\label{couple}
\begin{small}
\begin{tabular}{|l|l|}\hline
$\pi^0$& ${1\over \sqrt{2}}\bar p(\tilde D + \tilde F)p  - {1\over
\sqrt{2}}\bar n(\tilde D + \tilde F) n$\\&$ -{1\over \sqrt{2}}\bar
\Xi^0(\tilde D - \tilde F)\Xi^0 +{1\over \sqrt{2}}\bar
\Xi^-(\tilde D - \tilde F)\Xi^-$\\ & $+ \sqrt{2}
(\bar\Sigma^+\tilde F \Sigma^+ - \bar \Sigma^- \tilde F \Sigma^-)
+ \sqrt{2\over 3} (\bar \Sigma^0\tilde D \Lambda + \bar \Lambda
\tilde D \Sigma^0)$\\\hline

$\pi^+$& $\bar p(\tilde D + \tilde F) n + \bar \Xi^0(\tilde D -
\tilde F) \Xi^- $\\&$+ \sqrt{2} ( \bar \Sigma^0\tilde F \Sigma^- -
\bar \Sigma^+\tilde F \Sigma^0) + \sqrt{2\over 3} (\bar \Sigma^+
\tilde D \Lambda + \bar \Lambda\tilde D \Sigma^-)$\\\hline

$K^0$& $\bar p(\tilde D - \tilde F) \Sigma^+ - {1\over \sqrt{2}}
\bar n (\tilde D - \tilde F) \Sigma^0$\\&$ + \bar \Sigma^- (\tilde
D + \tilde F)\Xi^- - {1\over \sqrt{2}} \bar \Sigma^0(\tilde D +
\tilde F) \Xi^0)$\\ &$ - {1\over \sqrt{6} }\bar n (\tilde D + 3
\tilde F) \Lambda -{1\over \sqrt{6}}\bar \Lambda (\tilde D - 3
\tilde F) \Xi^0$\\\hline

$K^+$& $\bar n (\tilde D - \tilde F)\Sigma^- + {1\over \sqrt{2}}
\bar p(\tilde D -\tilde F) \Sigma^0 $\\&$+ \bar \Sigma^+ (\tilde D
+ \tilde F) \Xi^0 + {1\over \sqrt{2}} \bar \Sigma^0(\tilde D +
\tilde F) \Xi^-$\\&$ - {1\over \sqrt{6}} \bar p (\tilde D + 3
\tilde F) \Lambda - {1\over \sqrt{6}}\bar \Lambda (\tilde D - 3
\tilde F) \Xi^-$\\\hline

$\eta$& $-{1\over \sqrt{6}} \bar p ( \tilde D - 3 \tilde F) p -
{1\over \sqrt{6}}\bar n (\tilde D -3\tilde F) n $\\&$- {1\over
\sqrt{6}} \bar \Xi^0 (\tilde D + 3 \tilde F)\Xi^0 -{1\over
\sqrt{6}} \bar \Xi^-(\tilde D +3 \tilde F) \Xi^-$\\&$ +
\sqrt{2\over 3}
 (\bar \Sigma^-\tilde D \Sigma^- + \bar \Sigma^0 \tilde D \Sigma^0 + \bar
\Sigma^+\tilde D \Sigma^+) - \sqrt{2\over 3} \bar \Lambda \tilde D
\Lambda$
\\\hline
\end{tabular}
\end{small}
\end{table}

Since there is not an enhancement in  $\pi^0 p\bar p$ whose
amplitude is proportional to $\tilde F + \tilde D$, but there is
an enhancement in $K^+  \Lambda \bar p$ whose amplitude is
proportional to $3\tilde F + \tilde D$, there should be a
cancellation between $\tilde F$ and $\tilde D$ near threshold
which results in $\tilde F + \tilde D \approx 0$ and $\tilde F
\neq 0$. With this cancellation we have a consistent explanation
for observation and non-observation of threshold enhancement in
$J/\psi \to K^+ \Lambda \bar p$ and $J/\psi \to \pi^0 p\bar p$,
respectively.
%
It is clear that any other process whose amplitude is not
proportional to $\tilde F + \tilde D$ should show an enhancement
near threshold. In the $SU(3)$ limit, the relative strength of
enhancements near threshold in different channels are fixed if one
only keeps S-wave contributions for small $|\vec q|$. To show the
relative enhancement, we define the ratio:
\begin{equation}
R(MB'\bar B)= \frac{ \vert {\mathcal T}(J/\psi\to M B'\bar
B)\vert^2}
              {\vert {\mathcal T}(J/\psi\to K^+\Lambda \bar p)\vert^2},
\end{equation}
where spin summations are implied in the amplitude squared. In
experiment the amplitude squared can be measured, hence the
defined ratios. In the $SU(3)$ limit, the ratios $R(MB'\bar B)$
can be easily read off from Table I. For example for decays
involving a $K^+$, we have: $R(K^+ \Sigma^- \bar n):R(K^+ \Sigma^0
\bar p): R(K^+ \Lambda \bar p): R(K^+ \Xi^- \bar \Lambda)
=6:3:1:4$, for a given $\vert\vec q\vert^2$.

\par
In a realistic situation, there are $SU(3)$ breaking effects. For
example, a source of the breaking effects comes from the splitting
in meson masses and baryon masses. The enhancement in some
channels may be weakened by $SU(3)$ breaking effects. Some of them
are even forbidden kinematically, such as $J/\psi \to \eta \Xi^-
\bar \Xi^- (\Xi^0 \bar \Xi^0)$ although the corresponding matrix
elements are not zero. Part of $SU(3)$ breaking effects due to
mass splitting comes from $\vec p$ in the amplitude in
Eq.(\ref{sss}). To have some ideas about the mass splitting effect
on the threshold enhancements we neglect contributions from P-wave
or higher.  Then the matrix element squared is proportional to
$\vert \vec p\vert^2$. $\vert \vec p\vert^2$ varies in different
channels because of the mass differences. This factor will modify
the ratio. For example:
\begin{eqnarray}
&&R(K^0 \Sigma^+ \bar p)\nonumber\\
&& = { m^2_{\Lambda \bar p}\over
m^2_{\Sigma^+ \bar p }} {(M^2_{J/\psi}
  -m^2_{\Sigma^+\bar p} -m^2_{K^0} )^2 -4m^2_{\Sigma^+\bar p
  }m^2_{K^0}
  \over (M^2_{J/\psi}
  -m^2_{\Lambda \bar p} -m^2_{K^+} )^2  -4m^2_{\Lambda \bar p}m^2_{K^+}}
  \times 6.\nonumber
\end{eqnarray}
One should take this ratio as a function of $\vert\vec q\vert^2$
to compare with experimental data for small $\vert\vec q\vert^2$.
In this case $m_{\Lambda \bar p}$ and $m_{\Sigma^+ \bar p}$ are
the threshold masses $m_\Lambda + m_p$ and $m_{\Sigma^+} + m_p$.
Similarly one can obtain the ratios for other decays. We obtain
the non-zero ratios with: $R(\pi^0 \Xi^0 \bar \Xi^0): R(\pi^0
\Xi^- \bar \Xi^-): R(\pi^0 \Sigma^+ \bar \Sigma^+): R(\pi^0
\Sigma^- \bar \Sigma^-): R(\pi^0 \Lambda \bar
\Sigma^0):R(\pi^0\Sigma^0\bar \Lambda)
=0.54:0.51:1.50:1.42:0.63:0.63$, $R(\pi^+ \Xi^0 \bar \Xi^-):
R(\pi^+ \Sigma^0 \bar \Sigma^+): R(\pi^+ \Sigma^- \bar \Sigma^0):
R(\pi^+ \Lambda \bar \Sigma^+): R(\pi^+ \Sigma^- \bar
\Lambda)=1.04:1.48:1.44:0.63:0.62$, $R(K^0 \Sigma^+ \bar p): R(K^0
\Sigma^0 \bar n): R(K^0 \Lambda \bar n): R(K^0 \Xi^0 \bar
\Lambda)=4.71:2.32:0.99:0.77$, $R(K^+ \Sigma^- \bar n): R(K^+
\Sigma^0 \bar p): R(K^+ \Lambda \bar p): R(K^+ \Xi^- \bar
\Lambda)=4.59:2.35:1.:0.75$, $R(\eta p \bar p): R(\eta n \bar n):
R(\eta \Sigma^+ \bar \Sigma^+):R(\eta \Sigma^0 \bar
\Sigma^0):R(\eta \Sigma^- \bar \Sigma^-): R(\eta \Lambda \bar
\Lambda)= 6.28:6.24:0.22:0.21:0.19:0.48$.

From the above analysis one sees that there are many channels
where a near threshold enhancement can happen and in some of the
channels the enhancement is even larger than the enhancement in
$J/\psi \to K^+ \Lambda \bar p$.
A systematic
search for near threshold enhancement in all channels listed in
the above can reveal the detailed dynamics inducing the
enhancement.

 One can carry out a similar $SU(3)$ analysis for $J/\psi \to
\gamma B' \bar B$ decays by parameterizing the amplitudes as
\begin{eqnarray}
{\mathcal T} &=& Tr[ \bar B (
D_\gamma(\varepsilon_J,\varepsilon_\gamma, p, q) \{Q,
B\}\nonumber\\&& + F_\gamma(\varepsilon_J, \varepsilon_\gamma,
p,q) [Q, B])]\nonumber\\
&=&\bar p (F_\gamma + {1\over 3} D_\gamma) p + \bar \Sigma^+
(F_\gamma + {1\over 3} D_\gamma ) \Sigma^+ \nonumber\\&-& \bar
\Sigma^- (F_\gamma - {1\over 3} D_\gamma) \Sigma^- - \bar \Xi^-
(F_\gamma -
{1\over 3} D_\gamma) \Xi^-\nonumber\\
 &-&  {2\over 3}(\bar n
D_\gamma n + \bar \Xi^0 D_\gamma \Xi^0) -{1\over 3} \bar \Lambda
 D_\gamma \Lambda\nonumber\\ &+& {1\over 3} \bar \Sigma^0  D_\gamma
 \Sigma^0+ {1\over \sqrt{3}}(\bar \Lambda D_\gamma \Sigma^0   + \bar
\Sigma^0D_\gamma \Lambda) .
\end{eqnarray}
The general forms of $F_\gamma$ and $D_\gamma$ are given in eqs.
(\ref{gamma1}) and (\ref{gamma2}), respectively.

There may be cancellations among the $F_\gamma$ and $D_\gamma$
terms in certain channels, but not all of them. One therefore also
expects to observe enhancements in several other channels, if the
enhancement in $J/\psi \to \gamma p\bar p$ is confirmed. Since two
amplitudes are needed to specify the complete amplitudes, one
needs to measure another channel to fix the parameters.

In this work we have not dealt with the dynamical mechanism for the
enhancement in $J/\psi \to \gamma p \bar p$ and $J/\psi \to K^+
\Lambda \bar p$.
At present there is not enough information to decide the detailed
mechanism.
We emphasize, however, that the results obtained in this paper are
independent of any detailed dynamics for the enhancement effects.
Our predictions based on symmetry principles are therefore crucial
in further confirming the enhancement effects. Detailed study of
predictions in different channels will surely provide important
information about the enhancement effects.
 We strongly urge our experimental
colleagues to carry out systematic analysis about threshold
enhancement in $J/\psi \to \gamma B'\bar B$ and $J/\psi \to M
B'\bar B$.
\\

\noindent {\bf Acknowledgement}: HXG would like to thank Profs.
Han-Qing Zheng and Shi-Lin Zhu for useful discussions. HXG and MJP
would like to thank Prof. Chong-Xing Yue for hospitality at
Liaoning Normal University where part of the work was carried out.
This work was partly supported by grants from NSC and NNSFC.


\begin{thebibliography}{99}
\bibitem{bes1} J. Z. Bai et al., BES Collaboration, Phys. Rev.
Lett. {\bf 91}, 022001(2003).

\bibitem{bes2}M. Ablikim et al., BES Collaboration, e-print
hep-ex/0405050.

\bibitem{belle} K. Abe et al., Belle Collaboration, Phys. Rev.
Lett. {\bf 88}, 181803(2002); {\bf 89}, 151802(2002).

\bibitem{e1} A. Datta and O'Donnell, Phys. Lett. {\bf B
567}, 273(2003)

\bibitem{e2} X.A. Liu, {\it et al.}, e-Print hep-ph/0406118

\bibitem{ea} J. Rosner, Phys. Rev. {\bf D 68}, 014004 (2003)

\bibitem{e3} B.S. Zou and H. C. Chiang, Phys. Rev. {\bf D69}, 034004(2003)

\bibitem{e4} B. Kerbikov, A. Stavinsky and V. Fedotov, Phys. Rev. {\bf C69},
055205(2004)

\bibitem{e5} M.-L. Yan, S. Li, B. Wu and B.-Q. Ma, e-print hep-ph/0405087.

\bibitem{bugg} D.V. Bugg, eprint hep-ph/0406293.

\bibitem{zhu} C.-S. Gao and S.-L. Zhu, e-print hep-ph/0308205.

\bibitem{soni} C. Bernard et al.,
Phys.Rev. {\bf D32}, 2343(1985); C. Bernard, T. Draper, G.
Hockney, and A. Soni, Nucl.Phys.Proc.Suppl. {\bf 4}, 483(1988);
Also in Seillac Sympos.1987:0483 (QCD161:I715:1987)


\end{thebibliography}
\end{document}